# Simulation-Based Performance Evaluation of Routing Protocols in Vehicular Ad-hoc Network

**Mrs. Vaishali D. Khairnar**[*], **Dr. Ketan Kotecha**[**]

[*] Computer Engineering & Nirma University
[**] Computer Engineering & Nirma University

***Abstract-*** A Vehicular Ad-hoc Network (VANET) is a collection of wireless vehicle nodes forming a temporary network without using any centralized Road Side Unit (RSU). VANET protocols have to face high challenges due to dynamically changing topologies and symmetric links of networks. A suitable and effective routing mechanism helps to extend the successful deployment of vehicular ad-hoc networks. An attempt has been made to compare the performance of two On-demand reactive routing protocols namely AODV and DSR which works on gateway discovery algorithms and a geographical routing protocol namely GPSR which works on an algorithm constantly geographical based updates network topology information available to all nodes in VANETs for different scenarios. Comparison is made on the basis of different metrics like throughput, packet loss, packet delivery ratio and end-to-end delay using SUMO and NS2 simulator. In this paper we have taken different types of scenarios for simulation and then analysed the performance results.

## I. INTRODUCTION

VANET is autonomous and self-organizing wireless ad-hoc communication network. In this network vehicles are called nodes which involve themselves peer-to-peer for communication of information. This is new technology in India thus government has taken a huge attention on it. Many research projects related VANET are COMCAR [1], DRIVE [2], FleetNet [3] and NoW [4], CarTALK 2000 [5], CarNet [6]. Many different VANET applications such as Vehicle Collision Warning, Security Distance Warning, Driver Assistance, Cooperative Cruise Control, Dissemination of Road Information, Internet Access, Map Location, Automatic Parking and Driverless Vehicles. In this research paper we have analysed the performance of AODV DSR and GPSR routing protocol on CBR connection pattern with different pause time, speed time also different network parameters and different measured performance metrics such as Packet Delivery Ratio, Packet Loss, Throughput and End-to-End Delay of this three routing protocols are compared for their performance analysis.

## II. VEHICULAR AD-HOC NETWORK ROUTING PROTOCOLS

An ad-hoc routing protocol is a standard [9-10], that controls how vehicle nodes decide in which way to route the packets between computing device in vehicular ad-hoc network. There are different types of routing protocol in VANET such as proactive routing protocol, reactive routing protocol, hybrid routing protocol, topology based routing protocols and position based routing protocols. Existing unicast routing protocols of VANET is not capable to meet every traffic on highway road scenarios. They have also had some advantages and disadvantages. We have selected two reactive routing protocols i.e. AODV and DSR and one position-based routing protocol i.e. GPSR for simulation purpose analysis.

*Ad-hoc On Demand Distance Vector Routing Protocol (AODV)*

It is purely On-Demand route acquisition routing protocol. It is better protocol than DSDV network as the size of network may increase depending on the number of vehicle nodes [7] [12].

*Path Discovery Process [8] [12]*

In order to discover the path between source and destination, a route request message (RREQ) is broadcasted to all the neighbours who again continue to send the same to their neighbours, until the destination is reached. Every node maintains two counters: sequence number and broadcast-id in order to maintain loop-free and most recent route information. The broadcast-id is incremented for every RREQ the source node initiates. If an intermediate node receives the same copy of request, it discards it without routing it further. When a node forwards the RREQ message, it records the address of the neighbour from which it received the first copy of the broadcast packet, in order to maintain a reverse path to the source node. The RREQ packet contains: the source sequence number and the last destination sequence number know to the source. The source sequence number is used to maintain information about reverse route and destination sequence number tells about the actual distance to the final node.

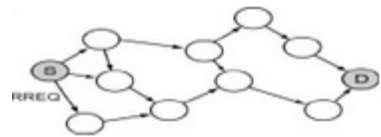

A)      *Source  node  S*

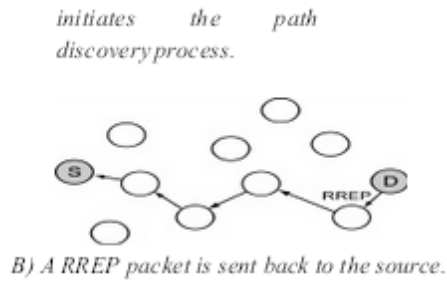

**Figure-1 AODV Path Discovery Process**

*Route Maintenance [12]*

A moving source node sends a new RREQ request packet to find a new route to the destination. But, if an intermediate node moves from its place, its upstream neighbour notices the move and sends a link failure notification message to each of its active upstream neighbours to inform them about the move until the source nodes is reached. After that the discovery process is again initiated.

III.    DYNAMIC SOURCE ROUTING PROTOCOL (DSR) [8] [12]

It is an On-Demand routing protocol in which the sequence of nodes through which a packet needs to travel is calculated and maintained as an information in packet header. Every mobile node in the network needs to maintain a route cache where it caches source routes that it has learned. When a packet is sent, the route-cache inside the node is compared with the actual route needs to be covered. If the result is positive, the packet is forwarded otherwise route discovery process is initiated again.

*A.  Route Discovery*

The source node broadcasts request-packets to all the neighbours in the network containing the address of the destination node, and a reply is sent back to the source node with the list of network-nodes through which it should propagate in the process. Sender initiates the route record as a list with a single element containing itself followed by the linking of its neighbour in that route. A request packet also contains an identification number called request-id, which is counter increased only when a new route request packet is being sent by the source node. To make sure that no loops occur during broadcast, the request is processed in the given order.

➢ If the pair (source address, request-id) is found in the list of recent route requests, the packet is discarded.
➢ If the host's address is already listed in the request's route record, then also the packet is discarded ensuring the removal of later copies of the same request that arrive by using a loop.
➢ When a destination address in the route request matches the host's address, a route reply packet is sent back to the source node containing a copy of this route.
➢ Otherwise, add this host's address to the route record field of the route request packet and rebroadcast the packet.

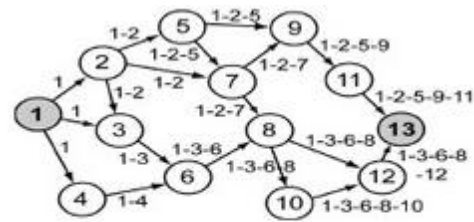

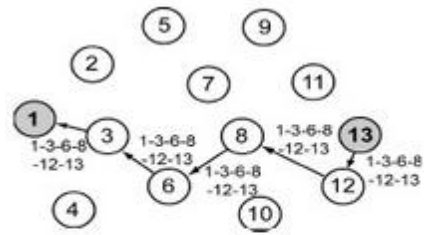

**Figure-2 DSR Route Discovery Process**

A route reply is obtained in DSR by two ways: Symmetric-links (bidirectional), in which the backward route is followed again to catch the source node. Asymmetric-links (unidirectional) needs to discover the route up to the source node in the same manner as the forward route is discovered.

*B.  Route Maintenance*

It can be accomplished by two ways: 1) Hop-by-Hop acknowledgement at the data link layer.
2) End-to-End acknowledgements.

The first method allows the early detection and retransmission of lost or corrupt packets in the data-link layer. If a transmission error occurs, a route error packet containing the address of node detecting the error and the host address is sent back to the sender. Whenever a node receives a route error packet, the hop in error is removed from the route cache and all routes containing this hop are truncated at that point. When the wireless transmission between two nodes does not work equally well in both directions, and then end-to-end replies on the application or transport layer may be used to indicate the status of the route from one host to the other.

IV.    GREEDY PERIMETER STATELESS ROUTING PROTOCOL (GPSR)

Greedy Perimeter Stateless Routing (GPSR) [25] is one of the best examples of position based routing. GPSR uses closest neighbours information of destination in order to forward packet. This method is also known as greedy forwarding. In GPSR each node has knowledge of its current physical position and also the neighbouring nodes. The knowledge about node positions provides better routing and also provides knowledge about the destination. On the other hand neighbouring nodes also assists to make forwarding decisions more correctly without the interference of topology information. All information about nodes position gathered through GPS devices. GPSR protocol normally devised in to two groups:

- ➢ Greedy forwarding: This is used to send data to the closest nodes to destination.
- ➢ Perimeter forwarding: This is used to such regions where there is no closer node to destination.

In other words we can say it is used where greedy forwarding fails. Further we will see in detail how these forwarding strategy works and what are issues in them.

### A. Greedy Forwarding

In this forwarding strategy data packets know the physical position of their destination. As the originator knows the position of its destination node so the greedy regions/hops are selected to forward the packets to the nodes that are closer to their destination. This process repeats until the packet successfully delivered to desired destination. Nearest neighbor's physical position is gathered by utilizing beaconing algorithms or simple beacons. When a neighboring node forwards packet to closer region to destination, the forwarding node receive a beacon message that contain IP address and position information. Then it updates its information in the location table. If forwarding node does not receive beacon from its neighboring node within a specific time period, it assumes that either neighbor fails to forward packet to region closer to destination or neighbor's is not in its radio range. So it removes its entry from location table [25]. The major advantage of greedy forwarding is that it holds current physical position of forwarding node. Thus by using this strategy total distance to destination becomes less and packets can be transmitted in short time period. Besides its advantages there are few drawbacks of this strategy i.e. there are some topologies used in it that limits the packet to move to a specific range or distance from the destination. Furthermore, this strategy fails when there are no closer neighbours available to destination.

### B. Perimeter Forwarding

Perimeter forwarding is used where greedy forwarding fails. It means when there is no next hop closest neighbour to the destination is available then perimeter forwarding is used. Perimeter forwarding uses nodes in the void regions to forward packets towards destination. The perimeter forwarding used the right hand rule. In right hand rule [25], the voids regions are exploited by traversing the path in counter clockwise direction in order to reach at specific destination. When a packet forward by source node, it forwarded in counter clockwise direction including destination node until it again reached at the source node. According to this rule each node involved to forward packet around the void region and each edge that is traversed are called perimeter. Edges may cross when right hand rule finds perimeter that are enclosed in the void by utilizing heuristic approach [24]. Heuristic has some drawbacks besides it provides maximum reach ability to destination. The drawback is that it removes without consideration of those edges which are repeated and this may cause the network partitions. To avoid this drawback another strategy is adopted that is described below.

### C. Planarized Graph

When two or more edges cross each other in a single graph is called planar graph. Relative Neighbourhood Graph (RNG) and Gabriel Graph (GG) [25] are two types of planar graphs used to remove the crossing edges. Relative neighbourhood graph (RNG) is defined as, when two edges intersect with radio range of each other and share the same area. For example, x and y are the two edges that share the area of two vertices x and y. The edge x, y are removed by using RNG because another edge from x towards v is already available Figure-3. Gabriel Graph (GG) is used to remove only those crossing edges which are in between the shared area of two nodes having the same diameter as the other nodes have. Figure-4 depicts GG: shows that the midpoint diameter is less than the diameter of node x or node y. Thus the edge from the x, y cannot be removed. So there is less network disconnection in the GG as compared to RNG.

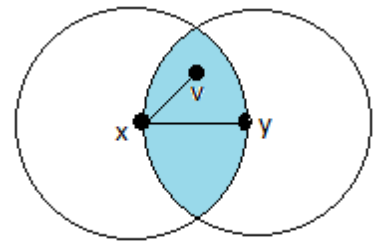

**Figure-3 Example of RNG**

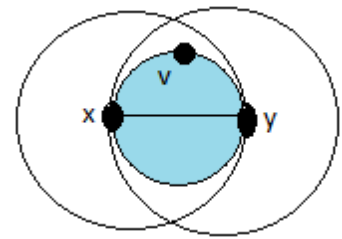

**Figure-4 Example of GG**

### D. Features of GPSR

GPSR combines the greedy forwarding with the perimeter forwarding to provide better routing decision on both full and Planarized network graph by maintaining neighbour's information in the location table. For the forwarding decisions in perimeter mode GPSR packet header include the following distinct characteristics [11].

- ➢ GPSR packet header has the flag identity that is used to identify whether packet is in greedy forwarding or in perimeter forwarding.
- ➢ It contains destination node physical address.
- ➢ GPSR packet header also contains location of packet in the perimeter mode and the location of the new face to take a decision whether to hold the packet in the perimeter mode or to return it to the greedy mode.
- ➢ GPSR also have the record of sender and receivers address of the packet when the edge's crosses in the new face.

GPSR also have several distinct characteristics that are if the packet is in perimeter mode then its location address is compared to forwarded node address and if distance to location and destination node is less then packet it switched to greedy mode to forward packet towards destination. GPSR discard those packets that are repeatedly forwarded as destination for such packets are not in range. The packets in perimeter mode never send twice through the same link if destination is in range. Overall GPSR is an efficient example of the position based routing that uses the geographic location of nodes and reduced usage of routing state on each node. Furthermore, it provides maximum robustness in highly dynamic wireless ad hoc networks.

*E. Issue in GPSR*

Besides GPSR certain characteristics, it suffers from several drawbacks. Greedy forwarding measured as unsuitable for the vehicular networks where the nodes are highly mobile and the node may not be able to maintain its next hop neighbours information as the other node may gone out of range due to high mobility. This can lead to data packets loss. The second problem may occur during beaconing mechanism that beacons may lost due to channel destruction or bad signal. This problem can lead to removal of neighbour information from location table [13]. GPSR uses Planarized graphs as its repair strategy where greedy forwarding fails. But these graphs perform well in the highway scenario due to their distributed algorithms [14]. These graphs does not perform well in such environment of vehicular communication where a lot of radio obstacles involves, in addition to this their distributed nature may lead to certain partition of network and may lead to packet delivery impossible. Hence there is need of such position based routing protocols, which merge position information with the road topological structure in order to make possible vehicular communication in presence of radio obstacles.

## V.  PROBLEM STATEMENT

The objective of the work is to compare the performance of the three routing protocols based on On-Demand Behavior, i.e. Ad-hoc On-Demand Distance Vector (AODV), Dynamic Source Routing (DSR) [8] [15] and Greedy Perimeter Stateless Routing (GPSR) protocols, for wireless ad-hoc networks based on the performance and comparison has been made on the basis of their properties like throughput, packet delivery ratio (PDR), end-to-end delay and data packet loss with respect to different scenarios - one by varying the number of nodes, again by varying the mobility of the nodes, again by varying the number of connecting nodes at a time and lastly by varying pause time.

The general objectives can be outlined as follows:
1) Study of Ad-hoc Networks
2) Get a general understanding of Vehicular Ad-hoc Networks
3) Study of different types of VANET Routing Protocol

4) Detailed study of AODV, DSR and GPSR
5) Generate a simulation environment that could be used for simulation of protocols
6) Simulate the protocols on the basis of different scenarios: by varying the number of nodes and by varying the traffic in the network
7) Discuss the result of the proposed work and concluding by providing the best routing protocol.

## VI.  METHODOLOGY

➢ Selection Techniques for Network Performance Evaluation

There are three techniques for performance evaluation, which are analytical modeling, simulation and measurement [12]. Simulation is performed in order to get the real-event results with no assumption as in case of analytical modelling.

➢ Random Waypoint Mobility Model

A node, after waiting a specified pause time moves with a speed between 0 km/h and $V_{max}$ km/h to the destination and waits again before choosing a new way point and speed.

## VII.  SIMULATION ASSUMPTIONS

The following assumptions are considered when building the TCL script [8][16-18]:
1) For simplicity, all flows in the system are assumed to have the same type of traffic source. Each sender has constant bit rate (CBR) traffic with the rate of data rate/number of stations packet per second.
2) The source node is fixed to 100 nodes with maximum connection is 60 nodes (to show a density condition) and if the nodes are varied for the calculation it is mentioned in area.
3) The implementation of grid and integrate between grid and routing protocols.

## VIII.  PERFORMANCE METRICS

*A.  Packet Delivery Ratio*

Packet delivery ratio is a very important factor to measure the performance of routing protocol in any network. The performance of the protocol depends on various parameters chosen for the simulation. The major parameters are packet size, no of nodes, transmission range and the structure of the network. The packet delivery ratio can be obtained from the total number of data packets arrived at destinations divided by the total data packets sent from sources. In other words Packet delivery ratio is the ratio of number of packets received at the destination to the number of packets sent from the source. The performance is better when packet delivery ratio is high. Mathematically it can be shown as equation (i).

Packet Delivery Ratio =  $\dfrac{\Sigma(\text{Total packets received by all destination node})}{\Sigma(\text{Total packets send by all source node})}$ ------------(i)

## B. Average End-to-End Delay

Average End-to-end delay is the time taken by a packet to route through the network from a source to its destination. The average end-to-end delay can be obtained computing the mean of end-to-end delay of all successfully delivered messages. Therefore, end–to-end delay partially depends on the packet delivery ratio. As the distance between source and destination increases, the probability of packet drop increases. The average end-to-end delay includes all possible delays in the network i.e. buffering route discovery latency, retransmission delays at the MAC, and propagation and transmission delay. Mathematically it can be shown as equation (ii).

$$D = \frac{1}{n} \, {}^{n}\Sigma_{i=1} \ (Tr_i - Ts_i) * 1000 \ [ms] ----------------------(ii)$$

Where
    D = Average E2E Delay
    i  = packet identifier
    $Tr_i$ = Reception time
    $Ts_i$ = Send time
    n  =  Number of packets successfully delivered

## C. Packet Loss

Packet Loss is the ratio of the number of packets that never reached the destination to the number of packets originated by the source. Mathematically it can be shown as equation (iii).
PL= (nSentPackets- nReceivedPackets)/ nSentPackets ------(iii)

Where
    nReceivedPackets = Number of received packets
    nSentPackets = Number of sent packets

## D. Packet Loss Ratio

Packet Loss Ratio is the ratio of the number of packets that never reached the destination to the number of packets originated by the source. Mathematically it can be shown as equation (iv).

    PLR     =     (nSentPackets-    nReceivedPackets)/ nSentPackets * 100 ----------------------(iv)

Where
    nReceivedPackets     =     Number of received packets
    nSentPackets     =     Number of sent packets

## E. Average Throughput

It is the average of the total throughput. It is also measured in packets per unit TIL. TIL is Time Interval Length. Mathematically it can be shown as equation (v).

    Average     Throughput     =     (recvdSize/(stopTime-startTime))*(8/1000) ----------------(v)

Where
    recvdSize     =     Store received packet's size
    stopTime     =     Simulation stop time
    startTime     =     Simulation start time

## IX.   SIMULATION RESULTS

Two On-Demand (Reactive) routing protocols namely Ad-hoc On-Demand Distance Vector Routing (AODV) and Dynamic Source Routing (DSR) and one Geographical (Position-Based) routing protocols namely Greedy Perimeter Stateless Routing (GPSR) protocols is used. The mobility model used is Random waypoint mobility model because it models the random movement of the vehicle mobile nodes.

**Scenario 1**: In this scenario, number of nodes connected in a network at a time is varied and thus varying the number of connections, through which the comparison graphs of AODV, DSR and GPSR, is obtained.

| Parameter | Value |
|---|---|
| Protocols | AODV, DSR, GPSR |
| Number of Nodes | 30, 50, 150, 300 |
| Simulation Time | 600 sec |
| Traffic Type | CBR |
| Routing protocol | AODV, DSR, GPSR |
| Transmission Range | 250 m |
| Mobility Model | Random Waypoint |
| Simulation area | 500 * 500 m |
| Node Speed | 20 m/s |
| Pause Time | 00 sec |
| Interface Type | Queue |
| Mac Protocol | 802.11Ext |
| Packet Size | 512 MB |
| Queue length | 50 |
| Radio Propagation Model | Two Ray Ground |

**Table-1: Various parameters used while varying number of connections**

| Varying Traffic | Packet Loss | Average E2E Delay | Packet Delivery Ratio | Average Throughput | Packet Loss Ratio |
|---|---|---|---|---|---|
| 30 | 248 | 120.442 | 93.6028 | 240.9 | 2.890 |
| 50 | 644 | 131.145 | 97.5757 | 278.97 | 4.908 |
| 150 | 799 | 130.306 | 98.1747 | 240.58 | 5.999 |
| 300 | 1285 | 129.825 | 92.3664 | 266.87 | 6.789 |

**Table-2.1 AODV**

| Varying Traffic | Packet Loss | Average E2E Delay | Packet Delivery Ratio | Average Throughput | Packet Loss Ratio |
|---|---|---|---|---|---|
| 30 | 246 | 127.754 | 72.8348 | 218.56 | 1.590 |
| 50 | 173 | 74.7002 | 45.1786 | 248.55 | 1.678 |
| 150 | 383 | 193.11 | 11.4177 | 190.18 | 1.909 |
| 300 | 313 | 142.524 | 1.2919 | 198.33 | 1.909 |

**Table- 2.2 DSR**

| Varying Traffic | Packet Loss | Average E2E Delay | Packet Delivery Ratio | Average Throughput | Packet Loss Ratio |
|---|---|---|---|---|---|
| 30 | 235 | 110.750 | 70.8090 | 210.56 | 1.050 |
| 50 | 160 | 70.7008 | 40.4567 | 214.55 | 1.150 |

| 150 | 280 | 110.90 | 10.990 | 150.90 | 1.190 |
| 300 | 280 | 90.00 | 1.989 | 140.89 | 1.190 |

Table- 2.3 GPSR

**Table- 2 Performance of AODV, DSR and GPSR with varying Number of Connections**

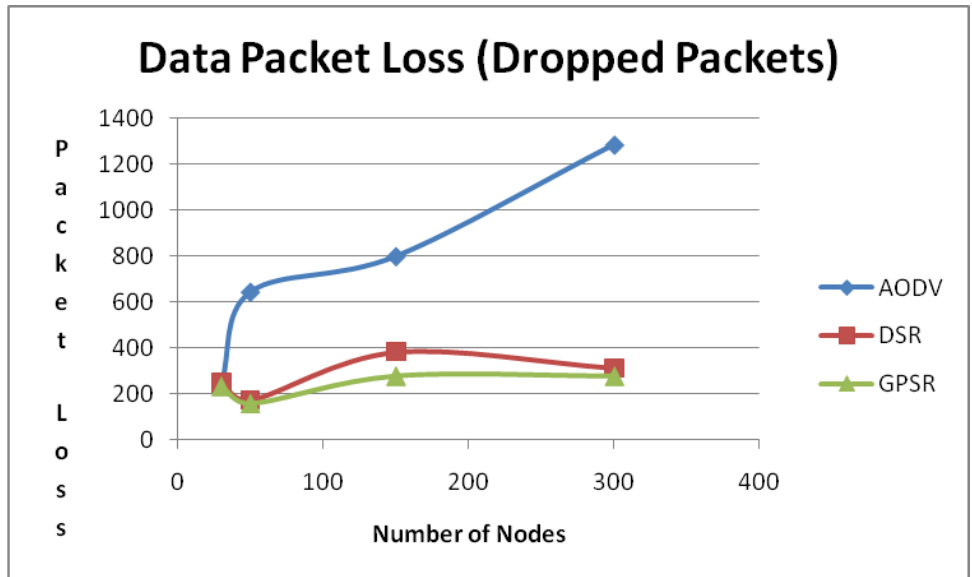

**Figure- 5 Data Packet Loss (Dropped Packets) for AODV, DSR and GPSR**

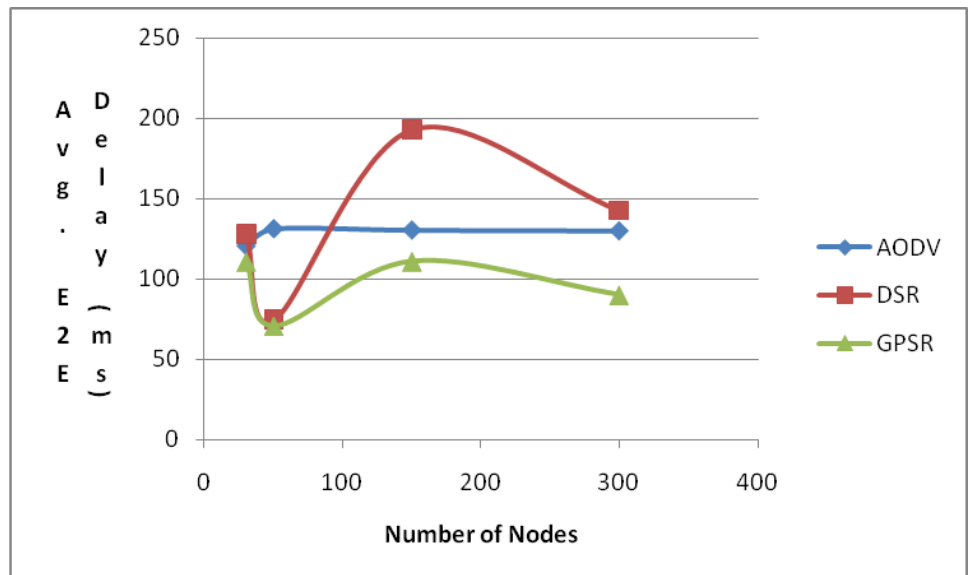

**Figure- 6 Average E2E Delay for AODV, DSR and GPSR**

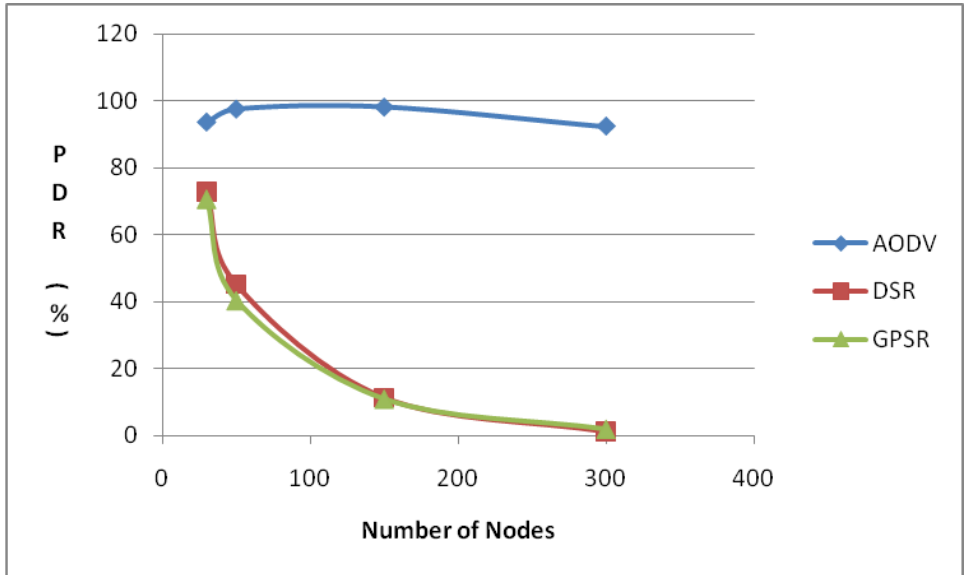

**Figure- 7 Packet Delivery Ratio for AODV, DSR and GPSR**

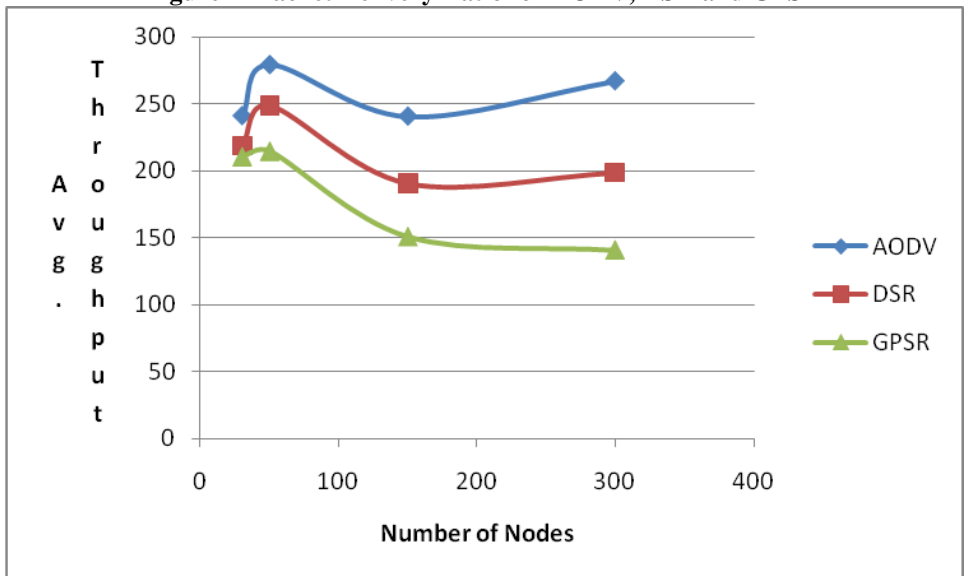

**Figure- 8 Average Throughput for AODV, DSR and GPSR**

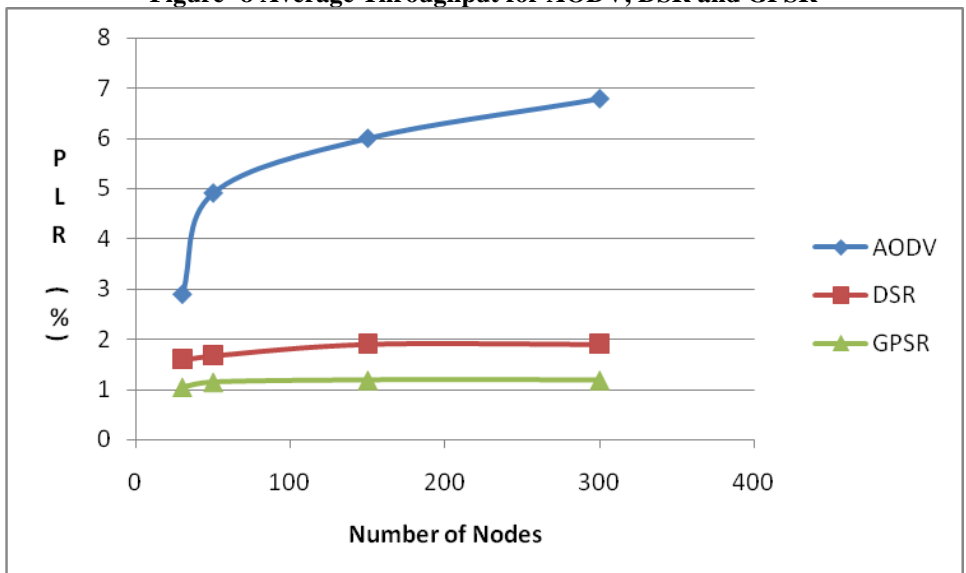

**Figure- 9 Packet Loss Ratio for AODV, DSR and GPSR**

**Scenario 2: Here in the second scenario the total number of vehicle nodes in the network at a time remains fixed and thus varying pause time of the network.**

| Parameter | Value |
|---|---|
| Protocols | AODV, DSR, GPSR |
| Number of Nodes | 200 with 100 connections |
| Simulation Time | 600 sec |
| Traffic Type | CBR |
| Routing protocol | AODV, DSR, GPSR |
| Transmission Range | 250 m |
| Mobility Model | Random Waypoint |
| Simulation area | 500 * 500 m |
| Node Speed | 10 m/s |
| Pause Time | 50 sec,100 sec, 150 sec, 200 sec, 250 sec, 300 sec. |
| Interface Type | Queue |
| Mac Protocol | 802.11Ext |
| Packet Size | 512 MB |
| Queue length | 64 |
| Radio Propagation Model | Two Ray Ground |

**Table-3: Various parameters used while varying pause time in the network**

| Pause Time | Packet Loss | Average E2E Delay | Packet Delivery Ratio | Average Throughput | Packet Loss Ratio |
|---|---|---|---|---|---|
| 50 | 1157 | 163.395 | 87.1369 | 204.97 | 2.890 |
| 100 | 995 | 104.604 | 92.892 | 452.67 | 2.345 |
| 150 | 1372 | 204.393 | 88.6116 | 248.94 | 2.134 |
| 200 | 1037 | 72.9835 | 92.1389 | 415.84 | 1.567 |
| 250 | 1355 | 101.22 | 95.859 | 608.61 | 1.456 |

**Table- 4.1 AODV**

| Pause Time | Packet Loss | Average E2E Delay | Packet Delivery Ratio | Average Throughput | Packet Loss Ratio |
|---|---|---|---|---|---|
| 50 | 541 | 140.519 | 2.96298 | 87.66 | 2.1890 |
| 100 | 754 | 227.774 | 6.31215 | 156 | 2.0981 |
| 150 | 1350 | 179.826 | 10.4053 | 117.3 | 1.8909 |
| 200 | 525 | 145.887 | 13.7914 | 221.97 | 1.7898 |
| 250 | 1434 | 208.651 | 35.0666 | 356.86 | 1.5678 |

**Table- 4.2 DSR**

| Pause Time | Packet Loss | Average E2E Delay | Packet Delivery Ratio | Average Throughput | Packet Loss Ratio |
|---|---|---|---|---|---|
| 50 | 450 | 130.908 | 2.45689 | 78.99 | 2.1345 |
| 100 | 680 | 234.900 | 2.56756 | 123 | 1.2347 |
| 150 | 590 | 139.080 | 8.76543 | 112.77 | 1.4568 |
| 200 | 300 | 123.879 | 9.78645 | 123.67 | 1.2349 |
| 250 | 560 | 178.094 | 15.6754 | 234.56 | 1.1230 |

**Table- 4.3 GPSR**
**Table- 4 Performance of AODV, DSR and GPSR with varying pause time in the network**

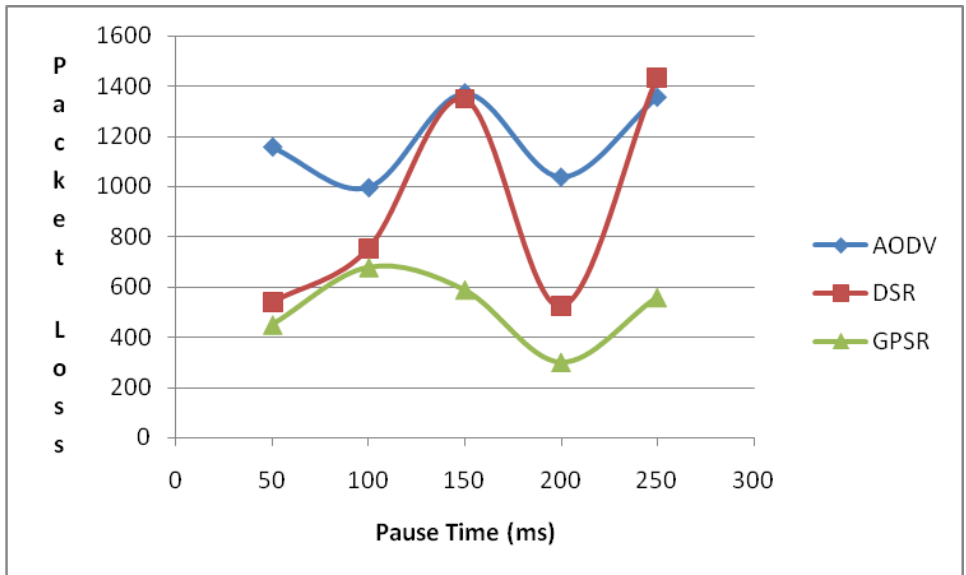

**Figure- 10 Data Packet Loss (Dropped Packets) for AODV, DSR and GPSR**

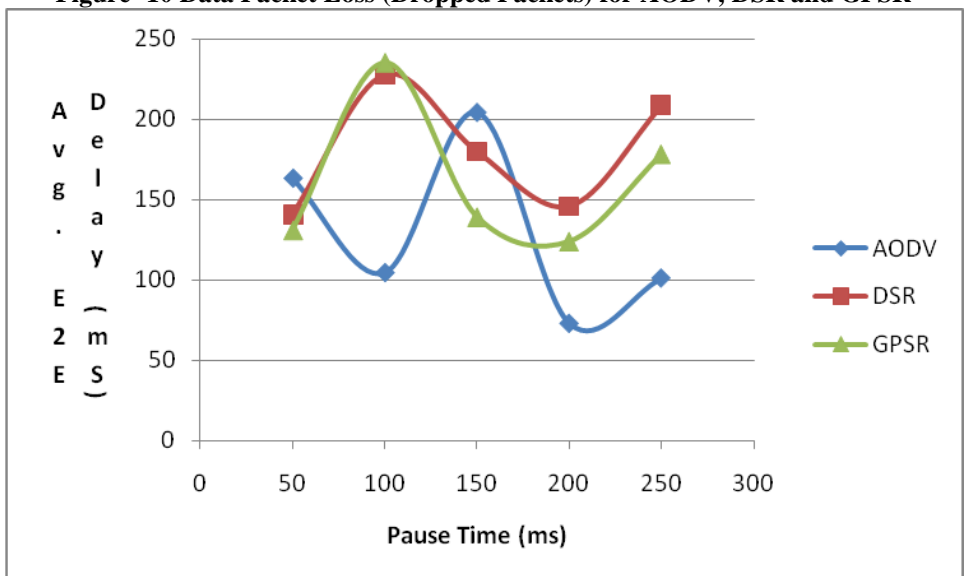

**Figure- 11 Average E2E Delay for AODV, DSR and GPSR**

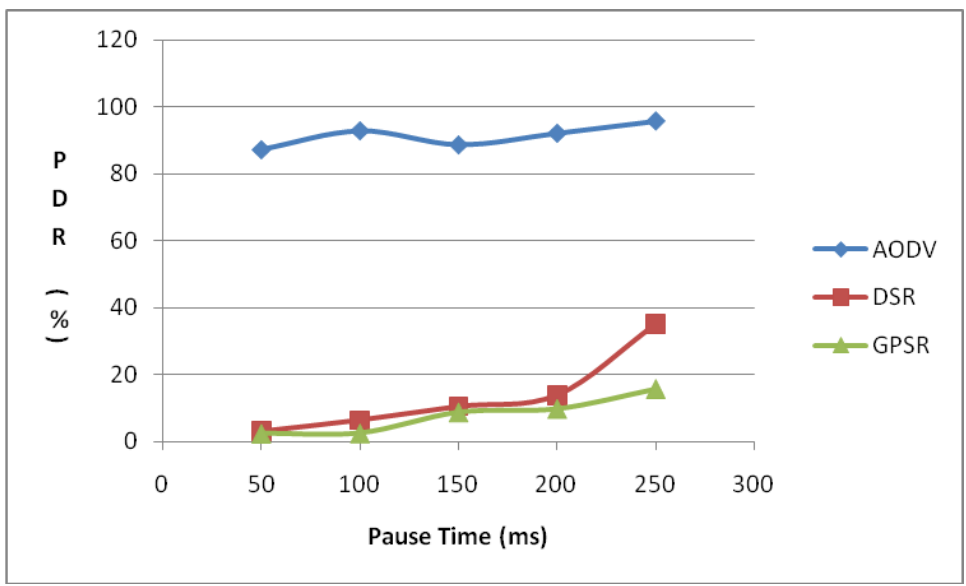

**Figure- 12 Packet Delivery Ratio for AODV, DSR and GPSR**

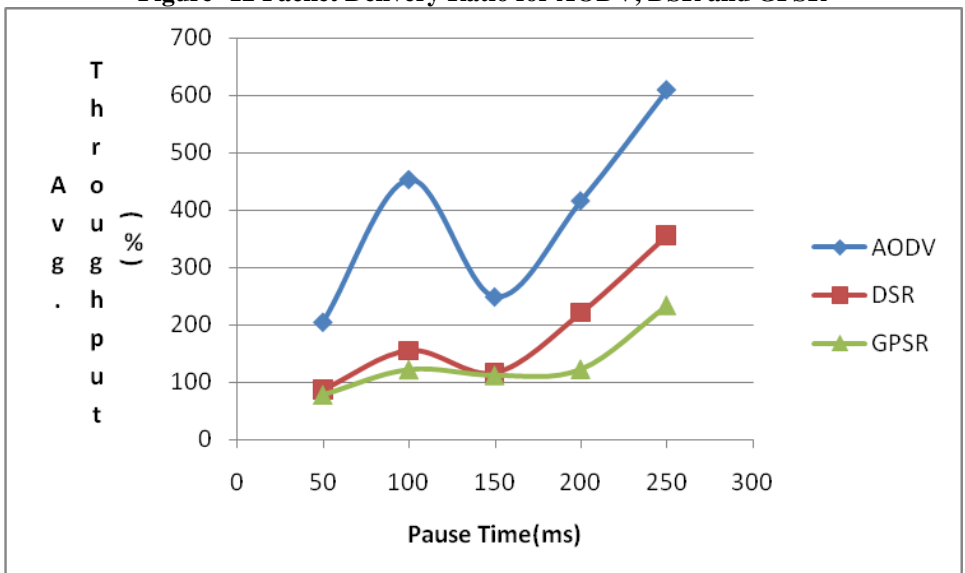

**Figure- 13 Average Throughput for AODV, DSR and GPSR**

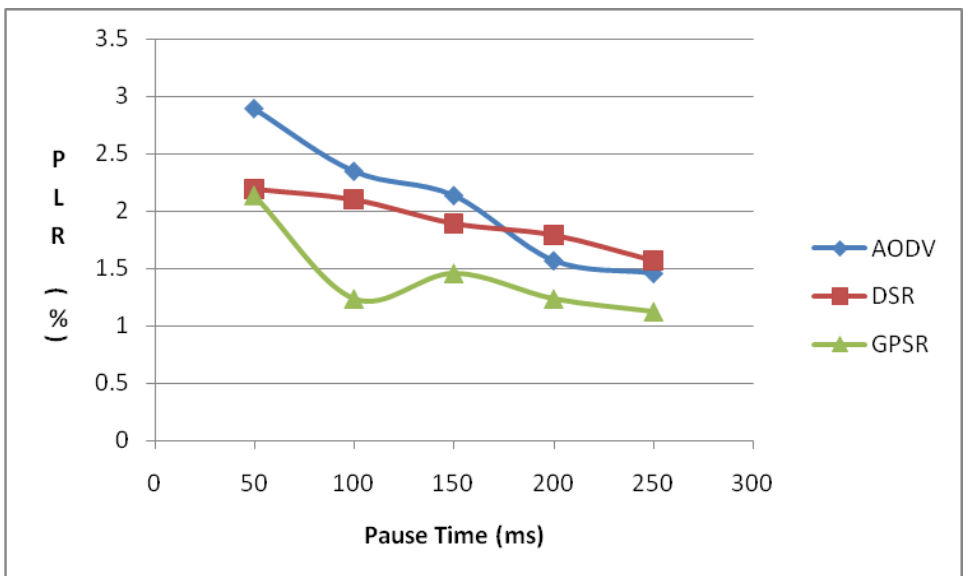

**Figure- 14 Packet Loss Ratio for AODV, DSR and GPSR**

**Scenario 3:** Here in the third scenario, the total number of vehicle nodes in the network at a time remains fixed and thus speed of the node with which they are moving in the area of 500 * 500 meter network.

| Parameter | Value |
|---|---|
| Protocols | AODV, DSR, GPSR |
| Number of Nodes | 200 with 100 connections |
| Simulation Time | 600 sec |
| Traffic Type | CBR |
| Routing protocol | AODV, DSR, GPSR |
| Transmission Range | 250 m |
| Mobility Model | Random Waypoint |
| Simulation area | 500 * 500 m |
| Node Speed | 10 m/s, 30 m/s, 50 m/s, 70 m/s, 90 m/s |
| Pause Time | 10 sec |
| Interface Type | Queue |
| Mac Protocol | 802.11Ext |

| Packet Size | 512 MB |
| --- | --- |
| Queue length | 50 |
| Radio Propagation Model | Two Ray Ground |

Table-5 various parameters used while varying mobility of the vehicle nodes i.e. speed of the nodes in the network

| Speed of the nodes | Packet Loss | Average E2E Delay | Packet Delivery Ratio | Average Throughput | Packet Loss Ratio |
| --- | --- | --- | --- | --- | --- |
| 10 | 1157 | 163.395 | 87.1639 | 204.87 | 2.5789 |
| 30 | 908 | 176.577 | 90.6245 | 249.17 | 10.678 |
| 50 | 954 | 323.638 | 88.1336 | 182.41 | 2.3456 |
| 70 | 1225 | 118.265 | 91.5398 | 327.57 | 2.4567 |
| 90 | 993 | 142.934 | 88.6138 | 217.87 | 1.5678 |

Table- 6.1 AODV

| Speed of the nodes | Packet Loss | Average E2E Delay | Packet Delivery Ratio | Average Throughput | Packet Loss Ratio |
| --- | --- | --- | --- | --- | --- |
| 10 | 541 | 140.519 | 2.95298 | 86.66 | 2.4567 |
| 30 | 127 | 159.535 | 0.18956 | 75.78 | 10.903 |
| 50 | 331 | 56.067 | 0.15583 | 52.18 | 1.2456 |
| 70 | 207 | 108.879 | 0.25082 | 53.29 | 2.3456 |
| 90 | 124 | 107.668 | 0.03373 | 11.02 | 2.5567 |

Table- 6.2 DSR

| Speed of the nodes | Packet Loss | Average E2E Delay | Packet Delivery Ratio | Average Throughput | Packet Loss Ratio |
| --- | --- | --- | --- | --- | --- |
| 10 | 528 | 135.900 | 1.8900 | 87.00 | 2.900 |
| 30 | 110 | 167.900 | 0.7829 | 72.00 | 11.780 |
| 50 | 135 | 78.900 | 0.6790 | 45.89 | 1.8902 |
| 70 | 178 | 108.890 | 0.1890 | 42.90 | 2.1900 |
| 90 | 109 | 107.099 | 0.1900 | 10.90 | 1.2899 |

Table- 6.3 GPSR

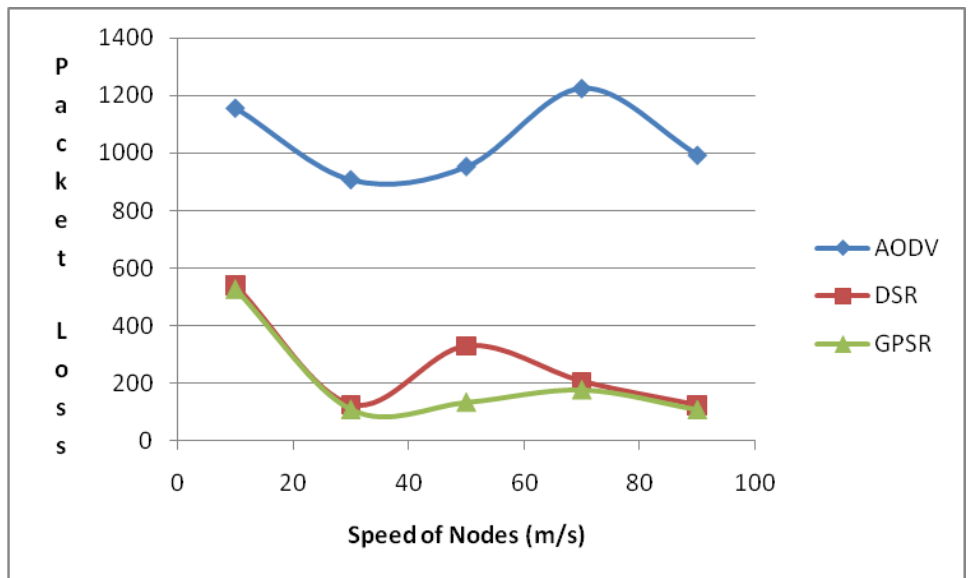

Figure- 15 Packet Loss (Dropped Packets) for AODV, DSR and GPSR

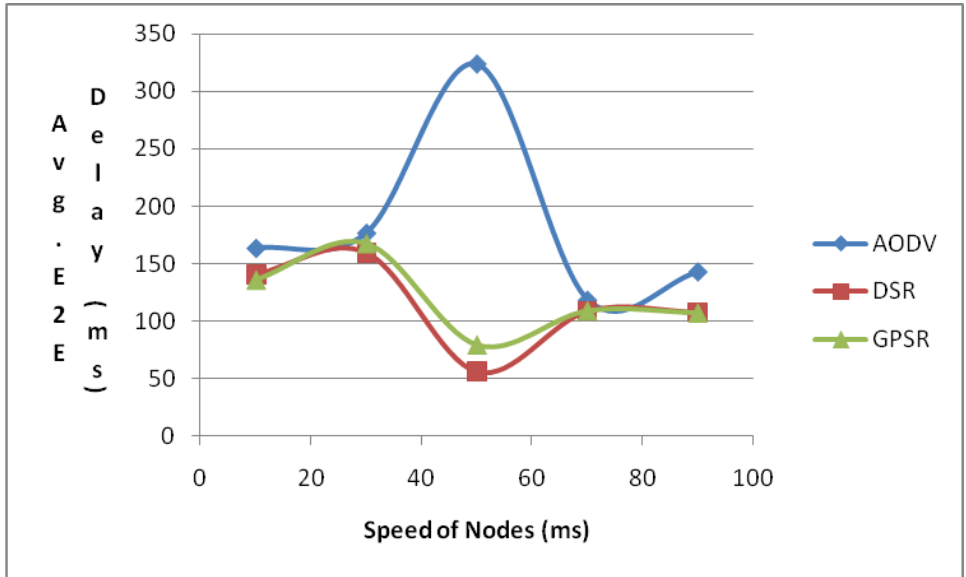

**Figure- 16 Average E2E Delay for AODV, DSR and GPSR**

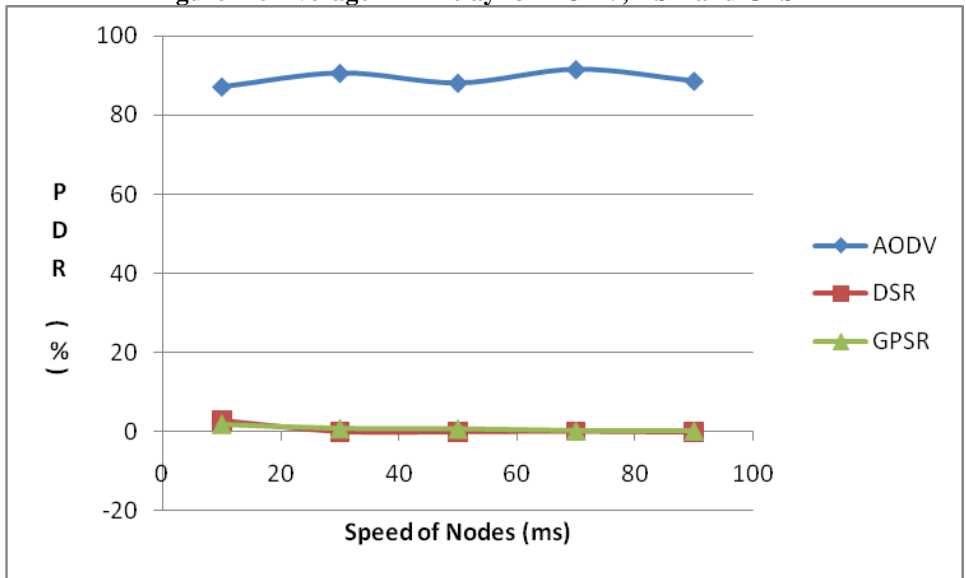

**Figure- 17 Packet Delivery Ratio for AODV, DSR and GPSR**

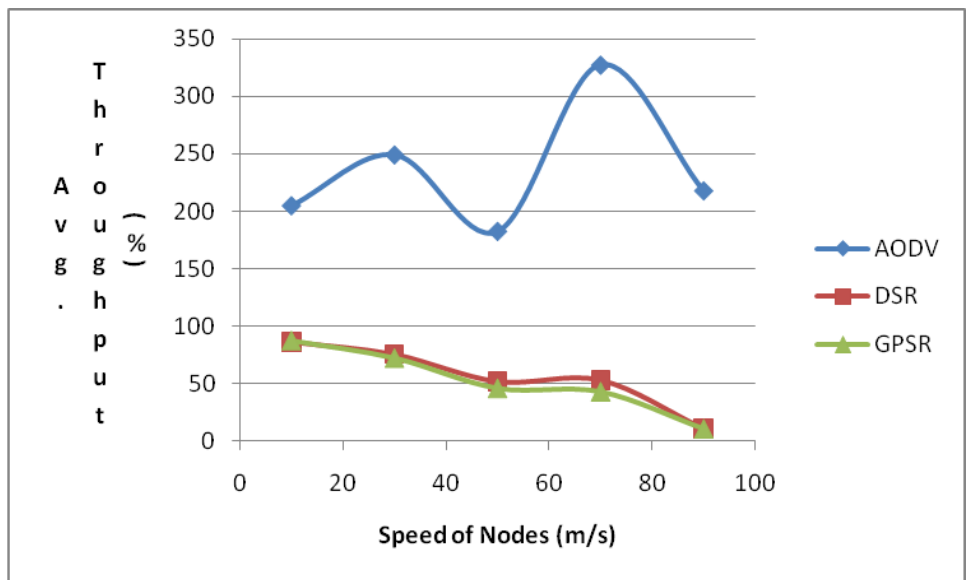

**Figure- 18 Average Throughput for AODV, DSR and GPSR**

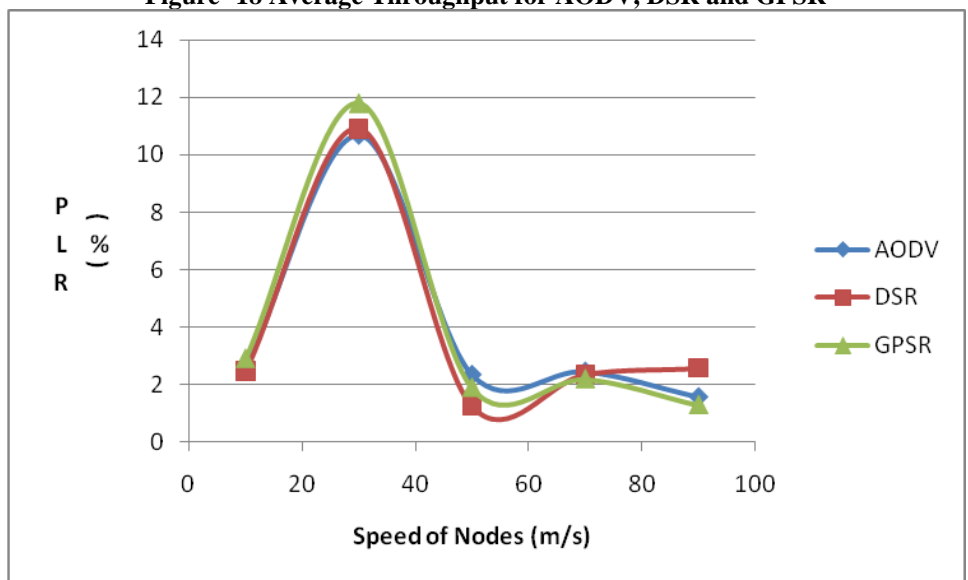

**Figure- 19 Packet Loss Ratio for AODV, DSR and GPSR**

## X.    ANALYSIS AND RESULTS

The paper shows the realistic comparisons of protocols which are both reactive and position based routing protocol and the simulation results agree based on theoretical analysis. The different scenarios were made in the SUMO and NS2.34. We run the simulation for 600 secs and generate the trace file from which we save the graphs for analysis and calculation as shown above. These graphs are found very helpful in the statistical analysis of these routing protocols performance. The required graphs were saved as the bitmap image for the statistical analysis.

**Scenario1:-** Number of Nodes Varied.

The first scenario is simulated and it generates the required trace file as shown in Figure-3. In this scenario, the vehicle nodes were simulated using Ad-hoc On Demand Distance Vector (AODV), Dynamic Source Routing Protocol (DSR) and GPSR routing protocol using CBR traffic application which were checked by different parameters such as E2E Delay, Packet Delivery Ratio, Packet Loss Ratio, Throughput etc. Graph show the Packet Delivery Ratio in percentage (%). The x-axis denotes the number of nodes and y-axis is PDR in %.

E2E Delay:- Performance of DSR increases and then decreases with increasing number of vehicle nodes, but the delay decreases with increasing number of vehicle nodes for GPSR network. For AODV, it varies with increasing number of vehicle nodes.

Packet Loss:- With increasing number of vehicle nodes AODV show worst-performance,  It remains same for all less number of vehicle nodes, but with increasing vehicle nodes AODV show maximum packet loss.

Packet Delivery Ratio:- Performance of AODV remains constant for increasing number of vehicle nodes , whereas performance of GPSR is more better than DSR and AODV.

Throughput:- The performance of AODV, DSR and GPSR remains almost constant for increasing number of vehicle nodes but GPSR and DSR shows better than AODV.

Packet Loss Ratio;- It remains same for all less number of vehicle nodes, but with increasing vehicle nodes AODV show maximum packet loss.

**Scenario 2**:- Pause Time Varied:-

E2E Delay:- AODV serves the best among all the protocols.
Packet Loss:- GPSR outperforms all other protocols in all conditions.

Packet Delivery Ratio:- GPSR performance better than AODV and DSR routing protocol.

Throughput:- GPSR outperforms the other two protocols but AODV shows better performance than DSR routing protocol.
Packet Loss Ratio:- GPSR outperforms all other protocols in all conditions.

**Scenario 3**:- Mobility of nodes is varied.

E2E Delay:- AODV performs constantly when speed of node changes whereas GPSR performs better than DSR.

Packet Loss:- GPSR and DSR performance better than AODV.

Packet Delivery Ratio:- DSR performs constantly in all conditions whereas AODV performs better than both GPSR and DSR.

Throughput:- DSR performance well in all conditions but GPSR performs better than AODV.
Packet Loss Ratio:- GPSR and DSR performance better than AODV.

## XI. CONCLUSION

AODV shows the best performance with its ability to maintain connection by periodic exchange of information required for TCP network. AODV performs best in case of packet delivery ratio and GPSR outperform others in case of throughput. Varying pause time, GPSR outperform others in case of packet loss and throughput, but overall AODV outperforms GPSR and DSR as in high mobility environment topology change rapidly and AODV can adapt to the changes, but with taking everything into account GPSR is better than others. At higher node mobility, AODV is worst in case of packet loss and throughput but performs best for packet delivery ratio, GPSR performs better than AODV for higher node mobility, in case of end-to-end and throughput but DSR performs best in case of packet loss. Hence, for real time traffic GPSR is preferred over DSR and AODV. Finally, from the above research work performance of AODV is considered best for Real-time and TCP network.

## AUTHORS

**First Author** – Mrs. Vaishali D. Khairnar, Computer Engineering & Nirma University
**Second Author** – Dr. Ketan Kotecha, Computer Engineering & Nirma University